\documentclass[pre, twocolumn, showpacs,amsmath,amssymb]{revtex4}

\usepackage{hyperref}
\usepackage{graphicx}
\usepackage{dcolumn}
\usepackage{xcolor}
\usepackage{braket}
\usepackage{commath}
\usepackage{adjustbox}
\usepackage{comment} 
\usepackage{ulem}
\usepackage{tikz}
\usetikzlibrary{shapes}

\newcommand{\rmd}{\mathrm{d}}
\newcommand{\bs}[1]{{\boldsymbol{#1}}}

\newcommand{\bq}{\bs{q}}
\newcommand{\bp}{\bs{p}}
\newcommand{\br}{\bs{r}}
\newcommand{\intdtwo}[1]{\int \frac{\rmd^3 #1}{(2\pi)^3}}% k-space integral d=3
\newcommand{\intdone}[1]{\int \frac{\rmd #1}{2\pi}}% k-space integral d=1
\newcommand{\intdr}[1]{\int \rmd^3 #1}

\begin{document}

\title{Induced dipole-dipole interactions in light diffusion from point dipoles}

\author{Nicolas Cherroret$^1$, Dominique Delande$^1$, Bart A. van Tiggelen$^2$}
\affiliation{$^1$Laboratoire Kastler Brossel, UPMC-Sorbonne Universit\'es, CNRS, ENS-PSL Research University, Coll\`ege de France; 4 Place Jussieu, 75005 Paris, France}
\affiliation{$^2$Universit\'e Grenoble Alpes, LPMMC, F-38000 Grenoble, France}

%\date{submitted version: April 25th, 2016}

\begin{abstract}
We develop a perturbative treatment of induced dipole-dipole interactions in the diffusive transport of electromagnetic waves through disordered atomic clouds. The approach is exact at order two in the atomic density and accounts for the vector character of light. It is applied to the calculation of the electromagnetic energy stored in the atomic cloud -- which modifies the energy transport velocity --  and of the light scattering and transport mean free paths. Results are compared to those obtained from a purely scalar model for light.
\end{abstract}


\maketitle

\section{Introduction}

Light propagating in thick cold atomic gases undergoes a multiple scattering process \cite{Houches99}. At the origin of this phenomenon, an incoming wave polarizes an atom, which re-emits a wavelet that can polarize another atom. If light travels over a distance much larger than the mean free path, this  elementary random process repeats itself many times so transport becomes diffusive on average. In this simple picture, the atomic scatterers seem to be independent from each other. This, however, may no longer be a good approximation when the number of atoms becomes large at the scale of the wavelength of the light \cite{Kaiser09}. Indeed, in this regime an atom that polarizes its neighbor can receive back the radiation, thus yielding an interaction energy between the two atoms. When considered from the point of view of the propagating wave, this phenomenon is referred to as dependent scattering. When considered from the point of view of the two atoms, it is known as induced dipole-dipole coupling (IDDC) and is, in particular, connected with the mechanisms of super and subradiance \cite{Dicke54, Tannoudji04}. 

Induced dipole-dipole coupling between pairs of scatterers affects the optical properties atomic clouds \cite{Keaveney12, Pellegrino14, Jennewein16}. In particular, in dilute clouds where light propagates by diffusion, they modify the diffusion coefficient. Accounting for these corrections is a highly nontrivial problem that requires to keep track of energy conservation (guaranteed by the Ward identity) in the perturbation theory. This task was accomplished in past work for scalar waves \cite{vanTiggelen94}. When dealing with multiple scattering of light however, an additional difficulty lies in the \emph{vector} nature of electromagnetic waves. Because of  this peculiarity, near-field effects are more pronounced than for scalar waves \cite{Naraghi15}, which may have strong consequences for the impact of IDDC on diffusion.
In this paper, we develop a multiple scattering theory of diffusive transport of electromagnetic (vector) waves through \emph{dilute} clouds of two-level atoms, treating in a rigorous way the cooperative interaction between pairs of scatterers. This allows us to derive the lowest-order dependent scattering corrections to the scattering and transport mean free paths and to the energy transport velocity, which are the three fundamental quantities governing light diffusion. We then compare these results to the previously studied scalar model \cite{vanTiggelen94} and comment on the differences. We finally discuss how our results could guide a description of multiple scattering of electromagnetic waves in atomic clouds of higher densities, where near-field effects were recently suggested to be responsible for the absence of Anderson localization \cite{Skipetrov14, Bellando14}. The main results of the paper are presented in Secs. \ref{sec_diffusion}, \ref{sec_vE}, \ref{sec_lstar} and \ref{sec_vector_scalar}. They are based on the transport theory for vector waves in random media, whose main lines are recalled in Appendix \ref{appendix_a}. Finally, some technical results are collected in Appendix \ref{appendix_b}.

\section{Diffusion of electromagnetic waves in atomic clouds}
\label{sec_diffusion}

Let consider a quasi-monochromatic electromagnetic wave of carrier frequency $\omega$ emitted by a point source located inside a three-dimensional, %isotropic, infinite disordered medium. We assume the latter to consist of a collection of two-level atoms of resonance frequency $\omega_0$ and density $n$, uniformly distributed in empty space. 
non-degenerate atomic gas of two-level atoms of resonance frequency $\omega_0$. For simplicity we assume the atomic transition to involve a non-degenerate ground state with angular momentum $J=0$ and an excited state with $J=1$. From here on we also neglect saturation effects as well as Doppler shifts resulting from the atomic motion. This reduces the model to a classical description of light scattering from uncorrelated point dipoles at rest. Since we consider a dilute atomic cloud, the number of atoms in an optical volume is typically small, namely
\begin{equation}
\label{diluteness}
\eta=\frac{4\pi n}{k^3}\ll 1,
\end{equation}
where $k=\omega/c$ is the wave number, $c$ the vacuum speed of light and $n$ the density of the atomic gas. Under this condition and in the hydrodynamic limit of long times and large distances from the source point, the disorder-averaged light intensity at time $t$ and point $\br$, scattered in the direction of the wave vector $\bp$ and detected in the polarization channel $\bs\epsilon$ is given by 
\begin{equation}
\overline{I}_\omega(\bp, \br,t)\!\sim\! \intdone{\Omega}\frac{\rmd^3 \bq}{(2\pi)^3}
\frac{A(\omega,p)e^{i\bq\cdot\br-i\Omega t}}{-i\Omega+D\bq^2}[1-(\hat{\bp}\cdot\bs\epsilon)^2],
\label{diff_intensity}
\end{equation}
where $\hat{\bp}=\bp/p$, $D$ is the diffusion coefficient and $A(\omega,p)$ is the spectral function (defined below). $\overline{I}_\omega(\bp, \br,t)$ is the optical analog of the Wigner distribution for massive particles. When integrated over $|\bp|$, it defines the so-called specific intensity \cite{Sheng95}. The term within the squared brackets signals the transverse character of light at large distances from the source point. A microscopic derivation of Eq. (\ref{diff_intensity}) is presented in appendix A, based on a semiclassical vector transport theory in random media initially developed in \cite{Barabanenkov95a, Barabanenkov95b}. Note that Eq. (\ref{diff_intensity}) implicitly assumes the existence of a diffusion pole at long times, which in three dimensions is a priori true only in the weak-disorder limit $k\ell^*\gg 1$, where $\ell^*$ is the transport mean free path of light. In dilute gases where Eq. (\ref{diluteness}) holds, this condition is however automatically fulfilled. Indeed, in the vicinity of the atomic resonance $k\ell^*=1/(n\sigma^*)\sim1/\eta\gg1$, where $\sigma^*$ is the resonant atomic cross-section \cite{Tannoudji04}. 

To  first order in $\eta$, the spectral function in Eq. (\ref{diff_intensity}) is given by
\begin{equation}
A(\omega,p)=\frac{2\omega}{\pi c^2}\frac{\omega/(v_\varphi\ell_s)}{(\omega^2/v_\varphi^2-p^2)^2+[\omega/(v_\varphi\ell_s)]^2},
\label{spectral_f}
\end{equation}
where $v_\varphi$ is the phase velocity, i.e. the speed of light divided by the effective refractive index of the atomic gas The explicit expression of $v_\varphi$ will be given below, see Eq. (\ref{vphi}). $\ell_s$ is the scattering mean free path, i.e the average distance traveled by light between two consecutive scattering events. In this paper, we will study $\ell_s$ by means of a second-order perturbation expansion in the parameter $\eta\ll1$. Using vector transport theory, we show in Appendix A that the diffusion coefficient of electromagnetic waves is given by
\begin{equation}
D=\frac{v_E\ell^*}{3},
\label{diff_coefficient}
\end{equation}
which is the same expression as for scalar waves \cite{vanTiggelen94}.
$v_E$ and $\ell^*$ are the two other fundamental transport quantities that we propose to study in this paper, up to second order in $\eta\ll1$.  The transport mean free path $\ell^*$ is the typical length scale for randomizing the direction of the wave vector \cite{Sheng95}. $v_E$ is the energy transport velocity, i.e. the speed of propagation of the average Poynting vector, and has been extensively studied theoretically \cite{Lagendijk93, vanTiggelen96, Cwilich92, Kogan92, Lubatsch05} and experimentally \cite{vanAlbada91, Labeyrie03, Sapienza07}.  As is well known, for resonant scatterers $v_E$ can be very \emph{different} from the phase velocity. Furthermore, when induced dipole-dipole interactions are considered, also  $\ell^*$ can be different from the scattering mean free path, sometimes used in the literature to characterize the diffusion coefficient \cite{Gero06, Gero07}.

\section{Energy transport velocity}
\label{sec_vE}

\subsection{Definition}

We start our analysis of IDDC by considering the energy transport velocity $v_E$, whose general formulation is provided by the transport theory for electromagnetic waves, recalled in Appendix \ref{appendix_a}:
\begin{equation}
v_E=\frac{c^2/v_\varphi}{1+a}.
\label{vE_def}
\end{equation}
In this relation, the phase velocity does not play a major role, unlike the parameter $a$ which significantly affects $v_E$, and on which we will focus on from here on. Physically, $a$ is the combined electromagnetic energy stored in the atomic dipoles and the interaction energy between them, relative to the electromagnetic energy in the surrounding environment \cite{Lagendijk93}. We show in Appendix A that up to second-order in $\eta$, $a$ is given by
\begin{eqnarray}
a&=&-\left(\frac{c}{\omega}\right)^2\left[\intdtwo{\bp}  \text{Im}\overline{G}^\perp(\omega,p)\right]^{-1}\nonumber\\
&&\times\text{Im}\intdtwo{\bp}
\overline{G}^\perp(\omega,p)\Sigma^\perp(\omega,p)+\mathcal{O}(\eta^3).
\label{adef}
\end{eqnarray}
Eq. (\ref{adef}) is similar to the corresponding expression for scalar waves given in \cite{Ozrin92}, except that the usual Green function is replaced by the transverse part $\overline{G}^\perp$ of the second-rank Green tensor $\overline{\bs{G}}$ that describes the average propagation of the electromagnetic field in the cold-atomic gas. $\overline{\bs{G}}$ obeys the Dyson equation \cite{Sheng95}
\begin{equation}
\overline{\bs{G}}=[\bs{G}_0^{-1}-\bs\Sigma]^{-1},
\label{Dysoneq}
\end{equation}
where $\bs{G}_0$ is the electromagnetic Green tensor in free space and $\bs\Sigma$ is the self-energy tensor. $\bs{\Sigma}$ features the elementary irreducible scattering processes from which multiple scattering sequences of the electromagnetic field are built on. As will be seen below, to order $\eta^2$, this includes both the process of light scattering from each individual atomic scatterer and the possibility for repeated scattering between pairs of atoms. The transverse component $\overline{G}^\perp(\omega,p)=[\omega^2/c^2-p^2-\Sigma^\perp(\omega,p)]^{-1}$ follows from the decomposition
\begin{equation}
\overline{\bs{G}}(\omega,\bp)=
\overline{G}^\perp(\omega,p)\bs P(\bp)+\overline{G}^\parallel(\omega,p)\bs Q(\bp),
\end{equation}
with a similar definition for $\Sigma^\perp(\omega,p)$. The tensors $\bs P(\bp)$ and $\bs Q(\bp)$ are the transverse and longitudinal projectors, respectively given by $P_{ij}(\bp)=\delta_{ij}-\hat{p}_i\hat{p}_j$ and $Q_{ij}(\bp)=\hat{p}_i\hat{p}_j$ in coordinate representation ($i,j=x,y,z$). %$\Sigma^\perp(\omega,p)$ is the transverse part of the two-rank self-energy tensor $\bs\Sigma(\omega,p)=\Sigma^\perp(\omega,p)\bs P(\bp)+\Sigma^\parallel(\omega,p)\bs Q(\bp)$, where $\bs P(\bp)=\bs{I}-\hat{\bp}\otimes\hat{\bp}$ and $\bs Q(\bp)=\hat{\bp}\otimes\hat{\bp}$ are the transverse and longitudinal projectors, respectively.

The fact that only the transverse parts of tensors $\bs\Sigma$ and $\overline{\bs{G}}$ appear in Eq. (\ref{adef}) is a consequence of the low-density approximation (\ref{diluteness}). Indeed, as discussed in Appendix A the longitudinal Green function $\overline{G}^\parallel(\omega,p)=[\omega^2/c^2-\Sigma^\parallel(\omega,p)]^{-1}$ does not contribute to $a$ at order 2 in density (note however that the longitudinal part of $\bs G_0$ does contribute to $\Sigma^\perp$, see below). We will come back to this point in Sec. \ref{localization_sec}.
%The explicit calculation of $a$ requires the knowledge of $\Sigma^\perp(\omega,p)$ and will be now presented in Sec. \ref{sec_results}. 

\subsection{Results}

\begin{figure}
\centering
\includegraphics[scale=0.58]{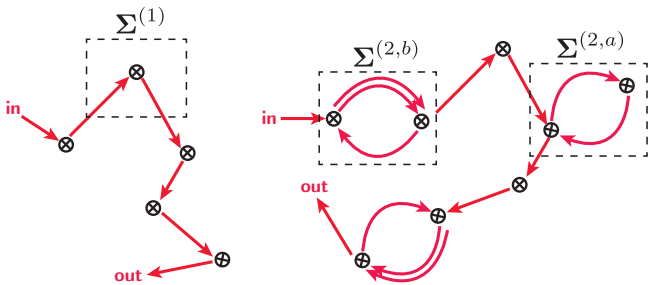}
\caption{(Color online) Sketch of light propagation in a dilute atomic cloud. Left: multiple scattering from independent atoms [$\bs\Sigma=\bs\Sigma^{(1)}$]: the propagating wave is never scattered more than one time by the same atom. Right: multiple scattering involving the possibility of repeated scattering (IDDC) between pairs of atoms [$\bs\Sigma=\bs\Sigma^{(1)}+\bs\Sigma^{(2)}$].
}
\label{scheme_propagation}
\end{figure}

\begin{figure}
\centering
\includegraphics[scale=0.5]{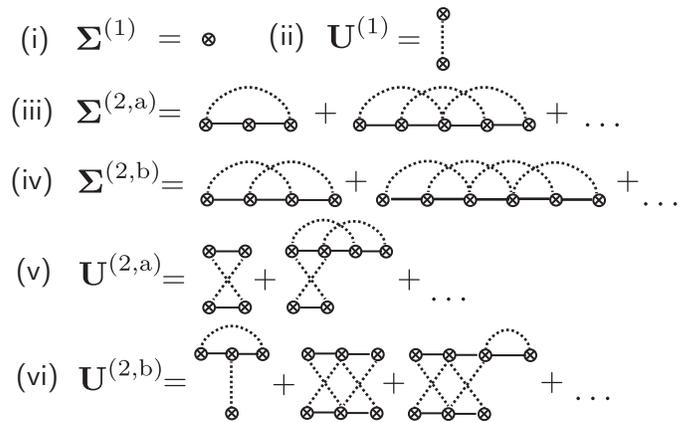}
\caption{First [(i), (ii)] and second-order [(iii), (iv), (v), (vi)] diagrams involved in the calculation of $\ell_s$, $a$ and $\ell^*$. Dotted arcs connect identical atoms. Solid lines refer to the free-space Green tensor $\bs G_0$. Circled crosses denote the atomic $t$ matrix.
}
\label{diagrams}
\end{figure}

Having expressed $a$ in terms of the fundamental irreducible tensor $\bs \Sigma$, we now explain how to evaluate this quantity. In order to capture the physics of IDDC, we make use of perturbation theory and expand $\bs \Sigma$ up to order $\eta^2$. Such an approach has been initially developed in \cite{vanTiggelen94} for scalar waves.  We here generalize it to vector waves and write
\begin{equation}
\bs\Sigma=\bs\Sigma^\text{(1)}+\bs\Sigma^\text{(2)}+\mathcal{O}(\eta^3).
\label{sigma_sum}
\end{equation}
When inserted into the Dyson equation (\ref{Dysoneq}), the first order of this expansion, $\bs\Sigma^\text{(1)}=\mathcal{O}(\eta)$, iterates a multiple scattering process where all atoms are independent, as illustrated in the left panel of Fig. \ref{scheme_propagation}. The self energy $\bs\Sigma^\text{(1)}$, depicted by a circled cross in Fig. \ref{diagrams}(i), is given by the $t$ matrix $t(\omega)$ of an individual two-level atom at frequency $\omega$, multiplied by the atomic density \cite{Tannoudji04, vanTiggelen94, deVries98}:
\begin{equation}
\bs{\Sigma}^\text{(1)}(\omega)=nt(\omega)\mathbf{1}=\frac{6\pi n}{k}\frac{\Gamma/2}{\delta+i\Gamma/2}\mathbf{1},
\label{sigma1}
\end{equation}
where $\mathbf{1}$ denotes the second-rank unit tensor and where we have introduced the natural width $\Gamma$ of the atomic transition and the detuning $\delta=\omega-\omega_0$ with respect to the resonance frequency $\omega_0$.
The second-order correction, $\bs\Sigma^\text{(2)}=\mathcal{O}(\eta^2)$, describes \emph{all} binary scattering processes \cite{footnote1}: in the course of the propagation, the light can be repeatedly scattered between two atoms, as illustrated in the right panel of Fig. \ref{scheme_propagation}. This phenomenon affects transport, and also implies a van der Waals type force between the two atoms of a pair.
The task of identifying all irreducible pair diagrams contributing to $\bs{\Sigma}^{(2)}$ has been accomplished in \cite{Tip90, vanTiggelen94}. The result can be recast as two infinite series $\bs{\Sigma}^{(2,a)}$ and $\bs{\Sigma}^{(2,b)}$ that are depicted in Fig. \ref{diagrams}(iii) and \ref{diagrams}(iv). $\bs{\Sigma}^{(2,a)}$ describes binary processes where the radiation incident on one atom eventually returns to the same one. It reads
\begin{equation}
\bs{\Sigma}^{(2,a)}(\omega)=\intdr{\br}\frac{n^2t^3\bs{G}_0^2(\br)}{\textbf{1}-t^2\bs{G}_0^2(\br)},
\label{sigma2a}
\end{equation}
where the frequency dependences of $\bs{G}_0$ and $t$ have been omitted to simplify the notations. In position space, the free Green tensor reads
\begin{eqnarray} 
\bs{G}_0(\br)&=&\left[-1+\frac{1}{ikr}+\frac{1}{(kr)^2}\right]\frac{e^{ikr}}{4\pi r}\bs{P}(\br)\nonumber\\
&&-2\left[\frac{1}{ikr}+\frac{1}{(kr)^2}\right]\frac{e^{ikr}}{4\pi r}\bs{Q}(\br)+\frac{\delta(\br)}{3k^2}\mathbf{1}.
\label{G0_eq}
\end{eqnarray}
Finally, the contribution $\bs{\Sigma}^{(2,b)}$ describes all processes where the radiation incident on one atom emerges from the second. It is given by
\begin{equation}
\bs{\Sigma}^{(2,b)}(\omega,p)=\intdr{\br}\frac{n^2t^4\bs{G}_0^3(\br)}{\mathbf{1}-t^2\bs{G}_0^2(\br)}e^{i\bp\cdot\br}
\label{sigma2b}
\end{equation}
and, unlike $\bs{\Sigma}^{(2,a)}$, displays a dependence on the wave number. Note that this series implicitly contains a local field correction $-n^2 t^2/3k^2$ stemming from the contact term in Eq. (\ref{G0_eq}), as was noted by Morice et al. \cite{Morice95}. This term is responsible for the so-called Lorentz-Lorenz correction to the atomic susceptibility in a dense medium, and  has no equivalent in the scalar model of light \cite{Ruostekoski97, Ruostekoski16}. If it were the only contribution to $\bs{\Sigma}^{(2)}$, it would shift the resonance line by the so-called Lorentz-Lorenz shift $\Delta\omega=-\pi n\Gamma/k^3$. In the present case, the other second-order contributions also affect the lineshape (see below).

\begin{figure}
\centering
\includegraphics[scale=0.52]{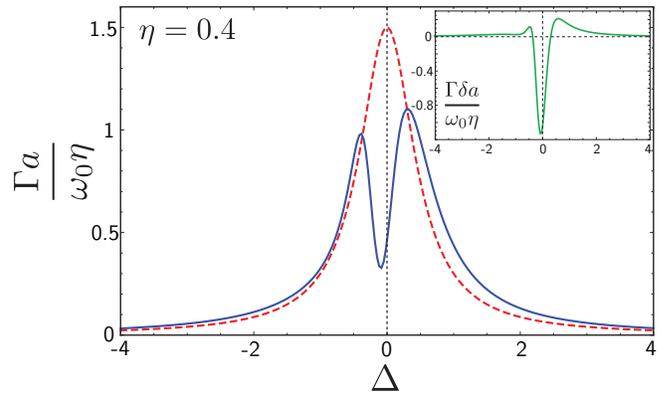}
\caption{(Color online) Main panel: stored electromagnetic energy per atom, $a/\eta$ [Eq. (\ref{a_tot})], in units of the quality factor $\omega_0/\Gamma$, for $\eta=0.4$ (solid blue curve). The dashed red curve is the independent-scattering approximation, Eq. (\ref{a_ISA}). Inset: Second-order contribution $\delta a$, Eq. (\ref{deltaa}).}
\label{velocity_vE}
\end{figure}

Making use of Eqs. (\ref{sigma_sum}), (\ref{sigma1}), (\ref{sigma2a}) and (\ref{sigma2b}), we can now evaluate Eq. (\ref{adef}) to order $\eta^2$. We find
\begin{equation}
a=a_\text{ISA}+\delta a,
\label{a_tot}
\end{equation}
where 
\begin{equation}
a_\text{ISA}=-\frac{nc}{\Gamma k}\text{Im}\, t
\label{a_ISA}
\end{equation}
and
\begin{equation}
\delta a=-\frac{n^2c}{\Gamma k^3}\text{Im} \frac{t^2}{4}-\frac{c}{\Gamma k}\text{Im}\, \Sigma^{\text{(2,a)}\perp}(\omega).
\label{deltaa}
\end{equation}
%\begin{equation}
%a=-\frac{3c^2}{2\Gamma\omega_0}\text{Im}\, nt
%-\frac{3c^4}{2\Gamma\omega_0^3}\text{Im} \frac{n^2t^2}{4}
%-\frac{3c^2}{2\Gamma\omega_0}\text{Im}\, \Sigma^{\text{(2,a)}\perp}.
%\label{explicit_a}
%\end{equation}
Let us briefly comment on these expressions. In the independent-scattering approximation (left panel in Fig. \ref{scheme_propagation}), $a\simeq a_\text{ISA}$ is the total electromagnetic energy stored in the individual atomic dipoles, relative to the electromagnetic energy in the surrounding environment. $a_\text{ISA}$ is shown in Fig. \ref{velocity_vE} as a dashed red curve as a function of the detuning normalized to the natural width of the transition, $\Delta=\delta/\Gamma$. We here assume a large quality factor, $\omega_0/\Gamma\gg 1$. In the vicinity of the resonance, $a_\text{ISA}\sim \eta \omega_0/\Gamma$ can be significantly larger than 1 even for a low density of scatterers. This phenomenon is responsible for the low velocity of light propagating through ensembles of resonant scatterers \cite{vanAlbada91, Labeyrie03, Sapienza07}. $\delta a$ contains two contributions. The first one [first term in the right-hand side of Eq. (\ref{deltaa})] is a trivial refractive index correction that originates from the renormalization of $k=\omega/c$ to $\omega/v_\varphi$ in $a_\text{ISA}$, where the phase velocity $v_\varphi$ is given by
\begin{equation}
v_\varphi=c\left[1+\frac{\text{Re}\Sigma^{(1)\perp}(\omega)}{2k^2}\right]+\mathcal{O}(\eta^2).
\label{vphi}
\end{equation}
In this formula, the term in the square brackets is the inverse of the refractive index of the cloud. It is here given only to lowest order, which is sufficient for the calculation of $a$ up to second order (second-order corrections to  the refractive index have been studied in \cite{Morice95}).
The second term in Eq. (\ref{deltaa}) involves $\Sigma^{\text{(2,a)}\perp}(\omega)$, the transverse component of Eq. (\ref{sigma2a}), and represents  the total interaction energy of the atomic pairs due to IDDC. Its explicit form is rather cumbersome and is given in Appendix \ref{appendix_b}. 
%Note that only $\Sigma^{(2,a)}$ appears in $\delta a_\text{DD}$, because the contribution (\ref{sigma2b}) vanishes when $\omega_0/\Gamma\to 1$ \cite{vanTiggelen94}, and we are left with
Note that when expanding Eq. (\ref{adef}) to second order in density and keeping only terms in lowest order in $\Gamma/\omega_0\ll 1$, one finds that the contribution of $\bs{\Sigma}^{(2,b)}$ vanishes. Thus, only the loop diagrams $\bs{\Sigma}^{(2,a)}$ contribute to $\delta a$, as is expected from the general expression of the potential that derives from a dispersion force \cite{Parsegian06, Bordag09}.

\begin{figure}
\centering
\includegraphics[scale=0.52]{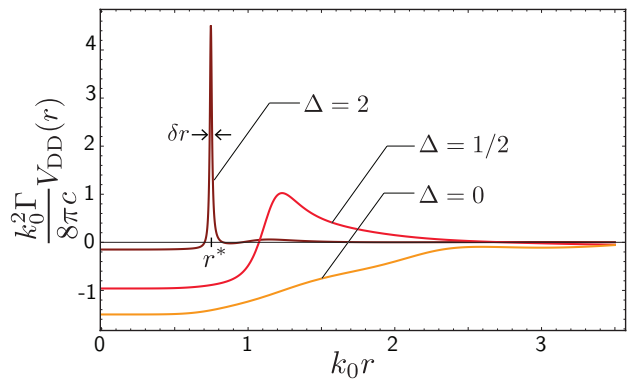}
\caption{(Color online) Induced dipole-dipole interaction potential $V_\text{DD}(r)$ between two atoms as a function of the inter-atomic distance $r$, for three values of $\Delta$. When $\Delta\gtrsim 1$, the curve displays a narrow subradiance peak. When $\Delta\to0$, this peak is smoothed out and the potential becomes essentially attractive.}
\label{VDD}
\end{figure}

We show the stored electromagnetic energy $a$ in Fig. \ref{velocity_vE} as a function of $\Delta$ for $\eta=0.4$ (blue curve). In the vicinity of the resonance, the curve displays a dip. This dip stems from IDDC, as is emphasized in the inset of Fig. \ref{velocity_vE}, which shows $\delta a$ as a function of $\Delta$: $\delta a$ is strongly negative around the resonance. In other words, the decrease of the energy transport velocity (\ref{vE_def}) is partially reduced as compared to the ideal situation where atoms are independent. To understand this phenomenon, it is instructive to look at the shape of the interaction potential $V_\text{DD}(r)$ between two atoms in a single pair near resonance \cite{Houches99, footnote2}:
\begin{equation}
V_\text{DD}(r)=-\frac{2c}{3\Gamma k_0}\text{Im}\text{Tr} \frac{t^3\bs{G}^2_0(\br)}{\bs{1}-t^2\bs{G}^2_0(\br)},
\end{equation}
where $k_0=\omega_0/c$. After summing $V_\text{DD}(r)$ over all pairs and integrating over $r$, one recovers the second term in the right-hand side of Eq. (\ref{deltaa}). The shape of $V_\text{DD}(r)$ is shown in Fig. \ref{VDD} for three positive values of $\Delta$ (the case $\Delta<0$ is similar). 
When $\Delta\gtrsim1$, it displays a narrow peak of width $\delta r\sim 1/(k_0\Delta^2)$ and centered at $r^*\sim 1/(k_0\Delta^{1/3})$. This peak corresponds to interatomic distances where light is resonant with the subradiant state that results from the coupling between the two atoms \cite{Dicke54} [a second very smooth peak (hardly visible in Fig. \ref{VDD}) corresponding to the superradiant state also shows up right next to the subradiant peak].
Far from resonance, $V_\text{DD}(r)$ is small everywhere except within the subradiance resonance, which is so peaked that it entirely controls the sign of $\delta a$ after integration over $r$. This explains the positive value of $\delta a$ in the wings of the resonance profile.
When $\Delta\ll1$ on the other hand, the subradiant peak is smoothed out so the near-field region where the potential is attractive extends over a broad range of interatomic distances. This makes $\delta a$ negative and explains the dip in Fig. \ref{velocity_vE}.

\section{Transport and scattering mean free paths}
\label{sec_lstar}

\subsection{Definition}

We now turn to the discussion of the transport mean free path $\ell^*$ for electromagnetic waves.
%The derivation of the transport mean free path for electromagnetic vector waves is presented in the Appendix for clarity. 
As shown in Appendix A, in the low-density limit (\ref{diluteness}), $\ell^*$ is given by
\begin{equation}
\frac{1}{\ell^*}=\frac{\langle(1-\hat{\bp}\cdot\hat{\bp}')U^{\perp}(\omega, k\hat\bp,k\hat\bp')\rangle_{\hat\bp'}}{8\pi}+\mathcal{O}(\eta^3).
\label{lstar_def}
\end{equation}
%\begin{equation}
%\frac{1}{\ell^*}=\left.\frac{\langle U^{\perp}_{\omega \bp\bp'}(1-\hat{\bp}\cdot\hat{\bp}')\rangle_{\hat{\bp}'}}{8\pi}\right|_{p=p'=k}+\mathcal{O}(\eta^3).
%\label{lstar_def}
%\end{equation}
where $\langle\ldots\rangle_{\hat\bp'}$ denotes the angular average over the direction of $\bp'$.
The fourth-rank tensor $\bs{U}$ involved in this formula is the irreducible \emph{intensity} vertex. $\bs{U}$ is to the average intensity what $\bs\Sigma$ is to the average field, and fulfills the Bethe-Salpeter equation \cite{Sheng95}
\begin{equation}
%\overline{\bs G\otimes\bs G}=\overline{\bs G}\otimes \overline{\bs G}^*+
%\overline{\bs G}\otimes \overline{\bs G}^*\cdot \bs U\cdot \overline{\bs G\otimes\bs G},
\overline{\bs G\otimes\bs G^*}=[(\overline{\bs G}\otimes \overline{\bs G}^*)^{-1}-\bs U]^{-1}.
\label{BSeq}
\end{equation}
As for $a$, at order $\eta^2$ only the transverse part $U^\perp$ of the intensity vertex appears in the definition of $\ell^*$. It is defined as $U^{\perp}(\omega,\bp,\bp')=\bs{P}(\bp)\cdot \bs U(\omega,\bp,\bp')\cdot \bs{P}(\bp')\equiv P_{ij}(\bp)U_{ij,kl}(\omega,\bp,\bp')P_{kl}(\bp')$ (summation over repeated indices is implied). 

The irreducible tensors $\bs U$ and $\bs\Sigma$ are not independent of each other. They are related through the Ward identity for electromagnetic waves, which guarantees energy conservation and is thus crucial for the global consistency of the perturbation theory. The full tensorial form of the Ward identity is given in Appendix A. It imposes the following relation between the transverse parts of $\bs U$ and $\bs\Sigma$: 
\begin{equation}
\frac{\langle U^{\perp}(\omega, k\hat\bp,k\hat\bp')\rangle_{\hat\bp'}}{8\pi}=-\frac{\text{Im} \Sigma^\perp(\omega,k)}{\omega/v_\varphi},
\label{Ward_transverse}
\end{equation}
where the phase velocity is given by Eq. (\ref{vphi}).
Making use of Eq. (\ref{Ward_transverse}), we rewrite Eq. (\ref{lstar_def}) under a form that will turn out to be more convenient for the perturbative expansion of the next section:
\begin{equation}
\frac{1}{\ell^*}=-\frac{\text{Im} \Sigma^\perp(\omega,k)}{\omega/v_\varphi}-\frac{\langle\hat{\bp}\cdot\hat{\bp}'U^{\perp}(\omega, k\hat\bp,k\hat\bp')\rangle_{\hat\bp'}}{8\pi}.
\label{lstar_def2}
\end{equation}
This definition of $\ell^*$ is exact at order $\eta^2$. The first term in the right-hand side defines the inverse of the scattering mean free path:
\begin{equation}
\frac{1}{\ell_s}=-\frac{\text{Im} \Sigma^\perp(\omega,k)}{\omega/v_\varphi}.
\label{ls_def}
\end{equation}
$\ell_s$ is the average distance traveled by light between two consecutive scattering events. It also gives the spatial decay rate of the average electromagnetic field in the disordered atomic cloud.

\subsection{Results}

\begin{figure}
\centering
\includegraphics[scale=0.51]{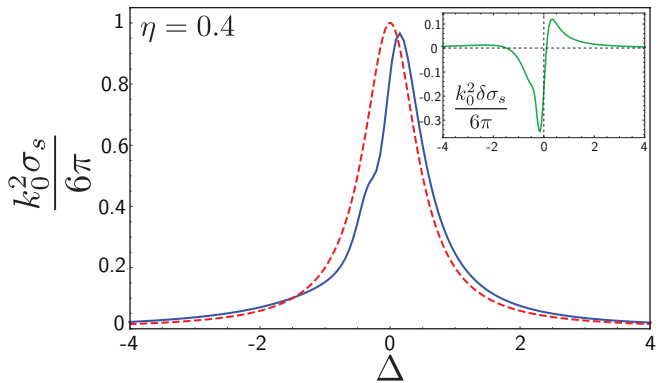}
\caption{(Color online) Main panel: scattering cross section, $\sigma_s$ [Eq. (\ref{sigma_s_tot})], in units of $6\pi/k_0^2$ ($k_0=\omega_0/c$), for $\eta=0.4$ (solid blue curve). The dashed red curve is the independent-scattering approximation, Eq. (\ref{sigma_ISA}). Inset: second-order contribution $\delta \sigma_s$, Eq. (\ref{delta_sigmas_vec}).
}
\label{ls_plot}
\end{figure}

Using the same perturbative expansion as in Sec. \ref{sec_vE}, we can straightforwardly evaluate the scattering mean free path $\ell_s$. We express the latter in terms of the scattering cross-section
\begin{equation}
\sigma_s\equiv \frac{1}{n\ell_s}=\sigma_\text{ISA}+\delta\sigma_s,
\label{sigma_s_tot}
\end{equation}
where
\begin{equation}
\sigma_\text{ISA}=-\frac{\text{Im}\, t}{k}.
\label{sigma_ISA}
\end{equation}
$\sigma_\text{ISA}$ is the  usual Lorentzian cross-section of an individual atomic dipole and is shown in Fig. \ref{ls_plot} as a function of $\Delta$ (dashed red curve).
The correction $\delta\sigma_s$ is given by
\begin{equation}
\delta\sigma_s=-\frac{n}{k^3}\text{Im}\frac{t^2}{4}-\frac{\text{Im}\Sigma^{(2,a)\perp}(\omega)+\text{Im}\Sigma^{(2,b)\perp}(\omega,k)}{nk}.
\label{delta_sigmas_vec}
\end{equation}
Again, beyond the independent-scattering approximation two types of corrections to the scattering cross-section show up. The first one [first term in the right-hand side of Eq. (\ref{delta_sigmas_vec})] is the refractive index correction to $\sigma_\text{ISA}$. The second correction [second term in the right-hand side of Eq. (\ref{delta_sigmas_vec})] is due to IDDC. It involves the transverse components of both the self-energies (\ref{sigma2a}) and (\ref{sigma2b}), whose explicit expressions are given in Appendix \ref{appendix_b}. 
$\sigma_s$ is shown in the main panel of  Fig. \ref{ls_plot} as a function of $\Delta$  (solid blue curve), and $\delta\sigma_s$ is shown in the inset. We see that the overall effect of second-order contribution is rather moderate.

According to Eq. (\ref{lstar_def2}), calculation of the transport mean free path requires the additional knowledge of the irreducible tensor $\bs{U}$. As for $\bs\Sigma$, we expand the latter as 
\begin{equation}
\bs{U}=\bs{U}^\text{(1)}+\bs{U}^\text{(2)}+\mathcal{O}(\eta^3).
\end{equation}
The first-order term, $\bs{U}^\text{(1)}=\mathcal{O}(\eta)$, is the well-known ladder vertex shown in Fig. \ref{diagrams}(ii) and given by $\bs{U}^\text{(1)}(\omega, \bp, \bp')=n |t(\omega)|^2\mathbf{1}$. Its contribution to $\sigma_s$ and $\sigma^*$ is already accounted in the first term in the right-hand side of Eq. (\ref{lstar_def2}), via the Ward identity (\ref{Ward_transverse}).
All second-order diagrams contributing to $\bs{U}^\text{(2)}$ have been identified in \cite{vanTiggelen94} in the scalar case. Among them, only the two types displayed in Fig. \ref{diagrams}(v) and \ref{diagrams}(vi) (as well as their complex conjugates, not shown in Fig. \ref{diagrams}) provide a non-vanishing contribution to the angular average  in Eq. (\ref{lstar_def2}). They are respectively given by
\begin{eqnarray}
&&\bs{U}^{(2,a)}(\omega, \bp, \bp')= \intdr{\br}\, n^2|t|^4 e^{i(\bp+\bp')\cdot\br}\nonumber\\
&&\times \frac{ \bs{G}_0(\br)\otimes \bs{G}_0^*(\br)}{[\mathbf{1}-t^2\bs{G}_0^2(\br)]\otimes[\mathbf{1}-t^2\bs{G}_0^2(\br)]^*}
\label{U2a}
\end{eqnarray}
and
\begin{eqnarray}
&&\bs{U}^{(2,b)}(\omega, \bp, \bp')= \intdr{\br}\, n^2|t|^2 e^{i(\bp-\bp')\cdot\br}\nonumber\\
&&\times\left\{\frac{\mathbf{1}}{[\mathbf{1}-t^2\bs{G}_0^2(\br)]\otimes[\mathbf{1}-t^{2}\bs{G}_0^{2}(\br)]^*}-\mathbf{1}\right\}.
\label{U2b}
\end{eqnarray}
Making use of Eqs. (\ref{sigma_s_tot}), (\ref{U2a}) and (\ref{U2b}), we can now evaluate the transport mean free path defined by Eq. (\ref{lstar_def2}). Expressing it in terms of the transport cross-section $\sigma^*$, we find
\begin{equation}
\sigma^*\equiv \frac{1}{n\ell^*}=\sigma_\text{ISA}+\delta\sigma^*,
\label{sigma_star_tot}
\end{equation}
where
\begin{eqnarray}
\delta\sigma^*=\delta\sigma_s-\frac{\langle \hat{\bp}\cdot\hat{\bp}' U^{(2,a)\perp}(\omega ,k\hat\bp, k\hat\bp')\rangle_{\hat\bp'}}{8\pi}\nonumber\\
-\frac{\langle \hat{\bp}\cdot\hat{\bp}' U^{(2,b)\perp}(\omega ,k\hat\bp, k\hat\bp')\rangle_{\hat\bp'}}{8\pi}.
\label{delta_sigma_vec}
\end{eqnarray}
\begin{figure}
\centering
\includegraphics[scale=0.51]{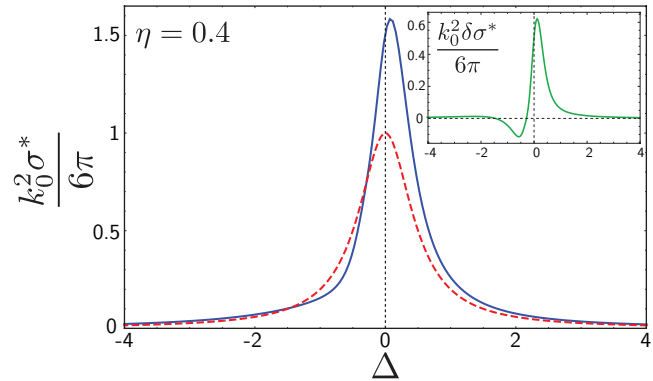}
\caption{(Color online) Main panel: transport cross section, $\sigma^*$ [Eq. (\ref{sigma_star_tot})], in units of $6\pi/k_0^2$, for $\eta=0.4$ (solid blue curve). The dashed red curve is the independent-scattering approximation, Eq. (\ref{sigma_ISA}). Inset: second-order contribution $\delta \sigma^*$, Eq. (\ref{delta_sigma_vec}).
}
\label{lstar_plot}
\end{figure}
The explicit expressions of the transverse components $U^{(2,a)\perp}$ and $U^{(2,b)\perp}$ are given in Appendix \ref{appendix_b}. $\delta\sigma^*$ is displayed in the inset of Fig. \ref{lstar_plot} as a function of $\Delta$, for $\eta=0.4$ (we again assume $\omega_0/\Gamma\gg 1$). We observe that IDDC brings essentially a positive correction to the transport cross-section (except in a narrow range on the red side of the transition). This is well visible in the main panel of Fig. \ref{lstar_plot}, which displays the full dependence of $\sigma^*$ on $\Delta$ (solid red curve). In other words, in the close vicinity of resonance and at low densities, the main effect of IDDC is to \emph{decrease} the transport mean free path of electromagnetic waves.
It is interesting to note that the presence of the last two terms in Eq. (\ref{delta_sigma_vec}) makes $\ell^*$ much more sensitive to IDDC than $\ell_s$.

\section{Vector versus scalar}
\label{sec_vector_scalar}

We finally compare the relative effect of IDDC  for vector and scalar waves. Mathematically, the essential difference lies in the near-field behavior of the Green function, which goes as $1/r^3$ for vector waves, see Eq. (\ref{G0_eq}), and in $1/r$ for scalar waves \cite{vanTiggelen94}. We anticipate that the manifestations of IDDC are more important for vector waves than for scalar waves, due to  the stronger weight on short distances. 

\subsection{Stored electromagnetic energy}

\begin{figure}
\centering
\includegraphics[scale=0.52]{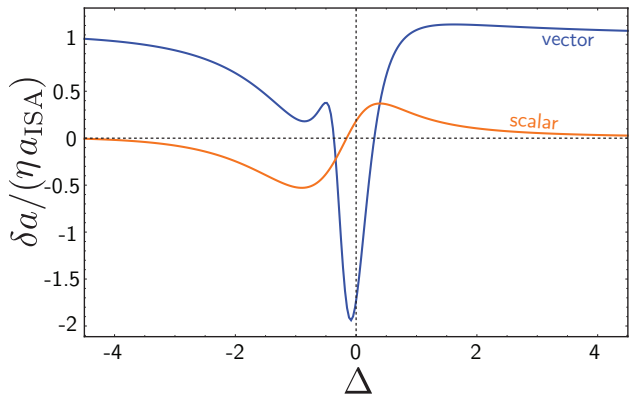}
\caption{(Color online) 
Relative correction $\delta a/(\eta a_\text{ISA})$ as a function of $\Delta$. The blue curve is the result for vector waves, Eq. (\ref{deltaa}), and the orange curve the result for scalar waves, Ref. \cite{vanTiggelen94}.}
\label{vE_vector_vs_scalar}
\end{figure}

We show in Fig. \ref{vE_vector_vs_scalar} the normalized correction $\delta a/(\eta a_\text{ISA})$ to the electromagnetic energy for vector (blue curve) and scalar (orange curve) waves as a function of $\Delta$ (up to a factor $\eta$, this quantity coincides with the first density correction to the dwell time for light in the scatterers \cite{vanTiggelen96}).
Since $a_\text{ISA}\propto\eta$ and $\delta a\propto\eta^2$, this ratio is independent of $\eta$. The shape of the two curves is markedly different both around resonance and away from it, which emphasizes the importance of near-field effects in the vector case. Near the resonance, no dip is visible in the scalar model, which is due to the absence of subradiance peak on the red side of the resonance for scalar waves. Far from the resonance, $\delta a/(\eta a_\text{ISA})$ does not fall to zero at large detuning for vector waves, unlike in the scalar model. This stems from the specific scaling of $\delta a$ with $\Delta$ when $\Delta\gg 1$:
\begin{equation}
\delta a\underset{|\Delta|\gg1}{\sim} \eta^2 \frac{\omega_0}{\Gamma}\frac{1}{\Delta^2}\sim \eta a_\text{ISA}\ \ \ \ \text{vector},
\label{large_Delta_a_vec}
\end{equation}
to be compared with the scalar result:
\begin{equation}
\delta a\underset{|\Delta|\gg1}{\sim} \eta^2\frac{\omega_0}{\Gamma}\frac{1}{\Delta^4}\sim {\eta}\frac{a_\text{ISA}}{\Delta^2}\ \ \ \ \text{scalar}.
\end{equation}
The scaling (\ref{large_Delta_a_vec}) is controlled by the subradiance peak which is very narrow when $|\Delta|\gg1$, see Fig. \ref{VDD} \cite{Svistunov88}.
Eq. (\ref{large_Delta_a_vec}) indicates that IDDC takes over the independent-scattering contribution at large detuning as soon as $\eta>1$ \cite{Nieuwenhuizen94, Sautenkov96}. In contrast, for scalar waves IDDC is completely negligible at large detuning even when $\eta\gtrsim 1$.
\begin{figure}
\centering
\includegraphics[scale=0.52]{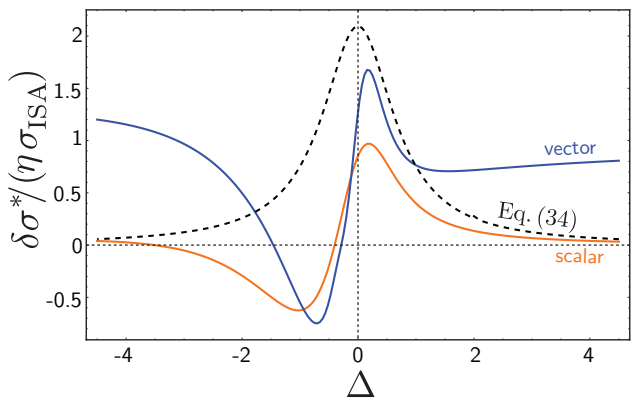}
\caption{(Color online) 
Relative correction $\delta\sigma^*/(\eta\sigma_\text{ISA})$ as a function of $\Delta$. The blue curve is the result for vector waves, Eq. (\ref{delta_sigma_vec}), and the orange curve the result for scalar waves, Ref. \cite{vanTiggelen94}. 
The dashed curve shows the contribution of the lowest-order crossed diagram only, calculated for scalar waves.}
\label{vector_vs_scalar}
\end{figure}

\subsection{Transport mean free path}

We also show in Fig. \ref{vector_vs_scalar} the normalized correction (independent of $\eta$) $\delta\sigma^*/(\eta\sigma_\text{ISA})$ to the transport cross section. Again, the results for scalar and vector waves differ at large detunings, for the same reason as for the stored electromagnetic energy. Note however that around the resonance, the change of $\sigma^*$ due to IDDC is qualitatively the same for scalar and vector waves, though more pronounced in the latter case.

Let us stress that \textit{all} scattering processes involved in light transport up to second-order in the atomic density are included in the perturbative approach discussed in this paper. Apart from the trivial refractive index correction to the independent scattering approximation, second-order corrections ensue from induced dipole-dipole coupling. Among all these binary processes, it is interesting to note that one is the familiar lowest-order crossed diagram [first digram in Fig. \ref{diagrams}(v)]. The latter has been argued to provide the leading-order density correction to $\sigma^*$ for scalar waves in continuous disordered potentials \cite{Kirkpatrick86, Eckert12}
and is given by
\begin{equation}
\frac{\delta\sigma^*_\text{Crossed}}{\sigma_\text{ISA}}=\frac{2\pi}{3}\frac{1}{k_0\ell_\text{ISA}},
\label{l_star_WL}
\end{equation}
where $\ell_\text{ISA}\equiv 1/(n\sigma_\text{ISA})$.
This contribution is shown in Fig. \ref{vector_vs_scalar} as a dashed black curve and, as expected, features a global decrease of $\ell^*$. By comparing with the exact second-order vector result that takes into account all IDDC processes however (blue curve), one clearly sees that Eq. (\ref{l_star_WL}) constitutes a poor approximation of $\delta\sigma^*/\sigma_\text{ISA}$. Even worse, for vector waves the lowest-order crossed diagram taken alone is in fact divergent. From these results, it thus appears that for light scattered from discrete objects like in dilute gases, the lowest-order crossed diagram contribution cannot be isolated from other  IDDC corrections.

\subsection{The question of localization}
\label{localization_sec}

We finally discuss the question of Anderson localization of light. According to Fig. \ref{vector_vs_scalar}, in the close vicinity of the atomic resonance,  IDDC tends to decrease slightly more $\ell^*$ in the vector case than in the scalar case. From this, one might be tempted to conclude that vector waves are at least as favorable as scalar waves for the observation of strong localization. This conclusion is however too naive, because it is not clear which role the near-field contributions discussed in this paper play in the regime $\eta\sim 1$ where localization might be expected. In fact, in the scalar case the description of strong localization is based on the study of the series of crossed diagrams \cite{Vollhardt80}. At low densities, this series is irrelevant in dimension 3 because it provides a (weak localization) contribution $\delta\sigma^*/\sigma_\text{ISA}\sim 1/(k_0\ell_\text{ISA})^2\propto \eta^2$, i.e. much smaller than the IDDC effects discussed in the present paper (which are of order $\eta$). A close inspection of the behavior of the series of crossed diagrams at $\eta\sim1$ might however be required to conclude on the fate of strong localization. To our knowledge, for vector waves such a task has not been accomplished yet. It is more challenging than in the scalar case 
%Presumably, the behavior of this series is different from the case of scalar waves
 for at least one reason: when $\eta\sim 1$, transport of vector waves can also be mediated by the longitudinal component $\overline{G}^\parallel$ of the Green function. The contribution of this mechanism to $\sigma^*$ has been estimated in \cite{Nieuwenhuizen94} in the dilute limit.  It was shown to be of third order and negative, $\delta\sigma^*/\sigma_\text{ISA}\sim-\eta^3<0$ \cite{footnote3}, thus possibly competing with localization at higher density. This could explain the absence of Anderson localization of light in atomic clouds predicted in recent work \cite{Skipetrov14, Bellando14}.

\section{Conclusion}

We have developed a diagrammatic perturbative treatment of binary induced dipole-dipole interactions for electromagnetic waves propagating in random ensembles of two-level atoms. As it describes all possible scattering processes at play up to second order in the density, our approach is rigorous and, in particular, fully satisfies the Ward identity. We have applied it to the analysis of the electromagnetic energy stored in the atomic gas and of the light transport mean free paths. In the close vicinity of the atomic resonance, both are decreased by IDDC. In particular, the stored energy displays a marked dip as a result of the attractive atomic interaction within pairs. This phenomenon is a genuine manifestation of near-field effects for vector waves and is absent for scalar waves.

An interesting question concerns the effect of IDDC on light transport at higher densities. In this regime, additional difficulties arise as the longitudinal component of the electromagnetic field can no longer be neglected in the kinetic equation for the light intensity. Longitudinal transport might also explain the recently predicted absence of Anderson localization \cite{Skipetrov14}. Given the elusive nature of three-dimensional Anderson localization of light in experiments \cite{Skipetrov16, Sperling16}, an analysis of this mechanism is undoubtedly an important challenge for future work.

\section{Acknowledgments}

The authors thank the Agence Nationale de la Recherche (grant ANR-14-CE26-0032 LOVE) for financial support. NC would like to thank Sergey Skipetrov, Robin Kaiser and Romain Gu\'erout for their comments and advice. DD thanks Chang-Chi Kwong and Romain Pierrat for useful discussions. BvT acknowledges Ad Lagendijk in earlier stages of this work. 

\appendix
\begin{widetext}
\section{Vector transport theory}
\label{appendix_a}

\subsection{Kinetic equation and Ward identity}

In this appendix, we present a transport theory for electromagnetic waves propagating in dilute atomic clouds, and use it to derive the diffusive solution (\ref{diff_intensity}) and formula (\ref{diff_coefficient}) for the diffusion coefficient, with $a$ and $\ell^*$ given by Eqs. (\ref{adef}) and (\ref{lstar_def}), respectively. As was shown in \cite{deVries98, vanTiggelen96}, for two-level atoms with a non-degenerate ground state, this problem can be equivalently tackled within a semi-classical formalism where atoms are modeled by dielectric point particles and light propagation is governed by the Helmholtz equation. This is the strategy we adopt here.

Let us so consider a quasi-monochromatic electromagnetic wave (spectral width $\Delta\omega$, carrier frequency $\omega\gg\Delta\omega$, polarization vector $\bs\epsilon_\text{in}$) emitted by a point source located inside a three-dimensional isotropic random medium. We assume the latter to consist of a collection of dielectric point scatterers uniformly distributed over space with density $n$. We describe them by an inhomogeneous relative dielectric function $\epsilon(\br)=\alpha_m\sum_i\delta(\br-\br_i)$, where the microscopic polarizability $\alpha_m$ depends on the atomic internal degrees of freedom ($\Gamma$ and $\omega_0$) \cite{deVries98}. The electromagnetic Green tensor $\bs G$ fulfills the Helmholtz equation
\begin{eqnarray}
-\bs\nabla\times\bs\nabla\times \bs{G}(\br',\br,\omega)&+&\frac{\omega^2}{c^2}\epsilon(\br)\bs{G}(\br',\br,\omega)=\delta(\br-\br')\mathbf{1}.
\end{eqnarray}
At a time $t\gg\Delta\omega^{-1}$, the disorder-averaged wave intensity at time $t$ and position $\br$ and detected in the polarization channel $\bs\epsilon_\text{out}$ and in the wave vector channel $\bp'$ is by definition
\begin{eqnarray}
\overline{I}_\omega(\bp', \br,t)=\intdone{\Omega}\intdtwo{\bq}\intdtwo{\bp}e^{i\bq\cdot\br-i\Omega t}
(\bs\epsilon_\text{in}\otimes
\bs\epsilon_\text{out}^*)\cdot\bs\Phi_{\omega\bp\bp'}(\bq,\Omega)\cdot
(\bs\epsilon_\text{in}^*\otimes \bs  \epsilon_\text{out}).
\label{intensity_Phi}
\end{eqnarray}
The intensity kernel $\Phi_{ij,kl}$ is a four-rank tensor related to the Green tensor through $\Phi_{ij,kl}=\overline{G_{ik}G_{jl}^*}$. Its momentum representation is explicitly given by
\begin{equation}
\bs\Phi_{\omega\bp\bp'}(\bq,\Omega)=
\overline{\langle \bp_+|\bs{G}(\omega_+)|\bp'_+\rangle\otimes\langle \bp'_-|\bs{G}^*(\omega_-)|\bp_-\rangle},
\end{equation}
where $\bp_\pm=\bp\pm\bq/2$, $\bp_\pm'=\bp'\pm\bq/2$, and $\omega_\pm=\omega\pm\Omega/2$. These conventions are summarized in Fig. \ref{tensor_conv}. In Eq. (\ref{intensity_Phi}), $\otimes$ denotes tensor product and the dots tensor contraction, with the same conventions as in \cite{Barabanenkov95a}.
\begin{figure}[h]
\centering
\includegraphics[scale=0.4]{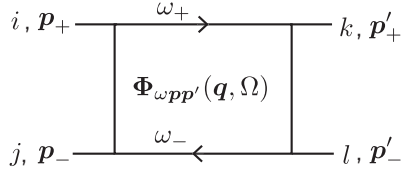}
\caption{Schematic representation of Eq. (\ref{intensity_Phi}), indicating the conventions for tensor indices and momenta. The upper line symbolizes $G_{ik}$ and the lower line $G^*_{jl}$.
\label{tensor_conv}}
\end{figure}

Given a wave of frequency $\omega$ coming from direction $\bp$, $\bs\Phi_{\omega\bp\bp'}(\br,t)$ can be interpreted as the average radiation density  at point $\br$ and time $t$, scattered in direction $\bp'$. $\bs\Phi_{\omega\bp\bp'}$ fulfills the tensorial Bethe-Salpeter equation (\ref{BSeq}). Combining the latter with the Dyson equation (\ref{Dysoneq}) for the average Green tensor, we find after a few algebraic manipulations \cite{Barabanenkov95a}:
\begin{eqnarray}
\left[\frac{i\Omega \omega}{c^2}\mathbf{1}-i\Delta\bs{L}_\bp(\bq)+\Delta\bs\Sigma_{\omega\bp}(\bq,\Omega)\right]
\cdot\bs\Phi_{\omega\bp\bp'}(\bq,\Omega)&=&
(2\pi)^3\delta(\bp-\bp')\Delta\bs{G}_{\omega\bp}(\bq,\Omega)\nonumber\\
&&+\intdtwo{\bp''}
\Delta\bs{G}_{\omega\bp}(\bq,\Omega)
\cdot\bs{U}_{\omega\bp\bp''}(\bq,\Omega)
\cdot\bs\Phi_{\omega\bp''\bp'}(\bq,\Omega).
\label{kinetic_eq}
\end{eqnarray}
All tensors that appear in this kinetic equation are of rank four. In particular, $\Delta\bs{G}_{\omega\bp}$ is defined as
\begin{equation}
\Delta\bs{G}_{\omega\bp}(\bq,\Omega)=\frac{1}{2i}
\left[\mathbf{1}\otimes\overline{\bs{G}}(\omega_+,\bp_+)-
\overline{\bs{G}}(\omega_-,\bp_-)\otimes \mathbf{1}
\right],
\end{equation}
where $\overline{\bs{G}}(\omega, \bp)=[k^2-\bs{L}(\bp)-\bs{\Sigma}]^{-1}$ with $\bs{L}(\bp)=\bp^2-\bp\otimes\bp$. $\Delta\bs\Sigma_{\omega\bp}$ has a similar definition, and
\begin{equation}
\Delta\bs{L}_{\bp}(\bq)=\frac{1}{2}
\left[\mathbf{1}\otimes\bs{L}(\bp_+)-
\bs{L}(\bp_-)\otimes \mathbf{1}
\right].
\end{equation}
Eq. (\ref{kinetic_eq}) is complemented by a conservation law, the Ward identity, which relates the irreducible vertices $\bs{U}$ and $\bs{\Sigma}$ \cite{Barabanenkov95a}:
\begin{equation}
\omega_-^2\mathbf{1}\otimes\bs\Sigma(\omega_+,\bp_+)
-\omega_+^2\bs\Sigma(\omega_+,\bp_+)\otimes\mathbf{1}=
\intdtwo{\bp'}\bs{U}_{\omega\bp\bp'}(\bq,\Omega)
\cdot \left[\omega_-^2\mathbf{1}\otimes\overline{\bs G}(\omega_+,\bp_+')
-\omega_+^2\overline{\bs G}(\omega_-,\bp_-')\otimes\mathbf{1}
\right].
\label{Ward_identity}
\end{equation}
%\begin{widetext}
%\begin{equation}
%-\frac{i\Omega}{\omega}
%\left[\bs{\Pi \Sigma}_{\omega\bp}(\bq)+\intdtwo{\bp'}\bs{\Pi G}_{\omega\bp'}(\bq)\bs{U}_{\omega\bp\bp'}(\bq,0)
%\right]=
%\Delta\bs\Sigma_{\omega\bp}(\bq,\Omega)-\intdtwo{\bp'}\Delta\bs{G}_{\omega\bp'}(\bq,\Omega)\bs{U}_{\omega\bp\bp'}(\bq,\Omega),
%\label{Ward_identity}
%\end{equation}
%\end{widetext}
%where
%\begin{eqnarray}
%\bs{\Pi \Sigma}_{\omega\bp}(\bq,\Omega)=
%\frac{\mathbf{1}\otimes\bs{G}(\omega,\bp_+)+
%\bs{G}(\omega,\bp_-)\otimes \mathbf{1}}{2}.
%\end{eqnarray}
Note the presence of the $\omega_\pm^2$ prefactors in Eq. (\ref{Ward_identity}), which are absent for matter waves obeying the Schr\"odinger equation \cite{Rammer98}. Here, they originate from the frequency dependence of the disorder ``potential'' $\omega^2\epsilon(\br)/c^2$ in the Helmholtz equation. As they depend on $\Omega$, these prefactors affect the dynamics of electromagnetic waves and ultimately give rise to the concept of energy transport velocity.

\subsection{Transverse fields approximation}

In this paper, we restrict ourselves to a second-order perturbation theory in density, based on the expansion of the irreducible tensors $\bs U$ and $\bs\Sigma$ up to order $\eta^2$, as explained in the main text. At order 2, the longitudinal component of the average Green tensor becomes irrelevant in the kinetic equation (\ref{kinetic_eq}) and the Ward identity (\ref{Ward_identity}) because it gives rise to terms of higher order in $\eta$ \cite{Nieuwenhuizen94}. 
%To illustrate this statement, let us consider for instance the term in the second line of Eq. (\ref{kinetic_eq}). To leading order, $\bs{U}_{\omega\bp\bp'}\propto\eta$, while the longitudinal component of $\int\rmd^3\bp'/(2\pi)^3 \Delta\bs G_{\omega\bp}$ is of the order of $\eta^2$. Consequently, $\int\rmd^3\bp'/(2\pi)^3\Delta\bs G_{\omega\bp}\cdot\bs{U}_{\omega\bp\bp'}\sim\eta^3$. 
For this reason, up to order $\eta^2$ it is sufficient to work with the \emph{transverse projection} of Eqs. (\ref{kinetic_eq}) and (\ref{Ward_identity}). This procedure is known as the ``transverse fields approximation'' and has been introduced in \cite{Barabanenkov95b, Barabanenkov97}. The projection is achieved by replacing every fourth-rank tensor $\bs{T}_{\omega\bp\bp'}(\bq,\Omega)$ in Eq. (\ref{kinetic_eq}) by
\begin{equation}
\bs{T}^\perp_{\omega\bp\bp'}(\bq,\Omega)=\bs{P}(\bp_+)\otimes\bs{P}(\bp_-)\cdot\bs{T}_{\omega\bp\bp'}(\bq,\Omega)\cdot\bs{P}(\bp'_+)\otimes\bs{P}(\bp'_-),
\end{equation}
where $\bs{P}(\bp)=\mathbf{1}-\hat{\bp}\otimes\hat{\bp}$. With this prescription, Eq. (\ref{kinetic_eq}) becomes
\begin{eqnarray}
\left[\frac{i\Omega \omega}{c^2}\mathbf{1}-i\bp\cdot\bq+\Delta\bs\Sigma^\perp_{\omega\bp}(\bq,\Omega)\right]\cdot\bs\Phi_{\omega\bp\bp'}^\perp(\bq,\Omega)&=&
(2\pi)^3\delta(\bp-\bp')\Delta \bs G_{\omega\bp}^\perp(\bq,\Omega)\nonumber\\
&&+\intdtwo{\bp''}
\Delta\bs{G}^\perp_{\omega\bp}(\bq,\Omega)
\cdot\bs{U}^\perp_{\omega\bp\bp''}(\bq,\Omega)
\cdot\bs\Phi^\perp_{\omega\bp''\bp'}(\bq,\Omega),
\label{kinetic_eq_trans}
\end{eqnarray}
with a similar projection for the Ward identity (\ref{Ward_identity}). Let us stress that while to order $\eta^2$ it is legitimate to neglect the longitudinal contributions to the kinetic equation, keeping them in the expression of $\bs\Sigma^\perp$ and $\bs{U}^\perp$ [via the longitudinal part of $\bs{G}_0$ in Eqs. (\ref{sigma2a}), (\ref{sigma2b}), (\ref{U2a}) and (\ref{U2b})] is on the other hand crucial. % for the correctness of the perturbation theory.

\subsection{Diffusive solution}

The general solution of Eq. (\ref{kinetic_eq_trans}) can be conveniently expressed in terms of a spectral decomposition of $\bs{\Phi}^\perp_{\omega\bp\bp'}$ originally introduced in \cite{Barabanenkov91, Barabanenkov95a}. In the limit of low frequencies and small wavenumbers ($\Omega\to 0$, $|\bq|\to0$) the behavior of $\bs{\Phi}_{\omega\bp\bp'}^\perp$ is governed by a single, second-rank transverse eigentensor $\bs\phi_{\omega\bp}$ with associated eigenvalue $\lambda$:
\begin{equation}
\Phi^{\perp ij,kl}_{\omega\bp\bp'}(\bq,\Omega)=\frac{\phi^{ik}_{\omega\bp}(\bq,\Omega)\phi^{jl}_{\omega\bp'}(\bq,\Omega)}{-i\Omega\omega/c^2+\lambda(\bq,\Omega)},
\label{diff_sol}
\end{equation}
where we have temporarily displayed tensor indices as superscripts. $\bs\phi_{\omega\bp}$ and $\lambda$ fulfill the eigenvalue equation
\begin{equation}
\left[\lambda(\bq,\Omega)
-i\bp\cdot\bq-\hat{\bs{  K}}_{\omega\bp}(\bq,\Omega)
\right]\cdot\bs\phi_{\omega\bp}(\bq,\Omega)=0,
\label{Kphi_eq}
\end{equation}
where we have introduced the fourth-rank tensor operator $\hat{\bs{  K}}_{\omega\bp}$ so that 
\begin{equation}
\hat{\bs{  K}}_{\omega\bp}(\bq,\Omega)\cdot\bs\phi_{\omega\bp}(\bq,\Omega)=
\intdtwo{\bp''}\left[
\Delta \bs G_{\omega\bp}^\perp(\bq,\Omega)\cdot\bs U^\perp_{\omega\bp\bp''}(\bq,\Omega)
-(2\pi)^3\delta(\bp-\bp'')\Delta\bs\Sigma^\perp_{\omega\bp''}(\bq,\Omega)
\right]
\cdot \bs{\phi}_{\omega\bp''}(\bq,\Omega).
\label{Kphi_eq2}
\end{equation}
The unknown quantities $\bs\phi_{\omega\bp}$ and $\lambda$ are determined from an expansion at small $\bq$ and $\Omega$. This is achieved by first expanding $\bs\phi_{\omega\bp}$ as
\begin{equation}
\phi_{ij}(\bq,\Omega)\sim \text{Im}\overline{G}^\perp(\omega,p)P_{ij}(\bp)+iq_k J_{ij,k}(\omega,\bp),
\label{expansion_phi}
\end{equation}
where we have introduced the third-rank current tensor $\bs{J}(\omega,\bp)$, yet to be determined. In this expansion, the proportionality of the term of zeroth order to $\text{Im}G^\perp(\omega,p)$ has been found by setting $\bq=0$ and $\Omega=0$ in Eq. (\ref{Kphi_eq}) and (\ref{Kphi_eq2}) and using that $\bs\phi_{\omega\bp}$ is a transverse tensor. Note that keeping an additional term of the order of $\Omega$ in Eq. (\ref{expansion_phi}) is not required here, as it would eventually gives a contribution of order $\Omega^2$ to $\bs{\Phi}^\perp_{\omega\bp\bp'}$. We also expand $\lambda(\bq,\Omega)$ as
\begin{equation}
\lambda(\bq,\Omega)\simeq\lambda(0,\Omega)+\delta\lambda(\bq,\Omega).
\label{expansion_lambda}
\end{equation}
Then, we expand the Ward identity and the kinetic equation to leading order in $\Omega$ and $\bq$, and combine them to obtain the following transport equation
\begin{eqnarray}
&&\left\{\frac{i\Omega \omega}{c^2}\left[\mathbf{1}+\bs\alpha^\perp(\omega,\bp)\right]-i\bp\cdot\bq\right\}\cdot\bs\Phi_{\omega\bp\bp'}^\perp(\bq,\Omega)=
(2\pi)^3\delta(\bp-\bp')\text{Im} \overline{G}^\perp(\omega,p)\bs{P}(\bp)\otimes\bs{P}(\bp)\nonumber\\
&&+\intdtwo{\bp''}
\bs{U}^\perp(\omega,\bp,\bp'')\cdot\left[
\text{Im}\overline{G}^\perp(\omega,p)
\bs{\Phi}^\perp_{\omega\bp''\bp'}(\bq,\Omega)-
\text{Im}\overline{G}^\perp(\omega,p'')
\bs{\Phi}^\perp_{\omega\bp\bp'}(\bq,\Omega)\right],
\label{kinetic_eq_tr}
\end{eqnarray}
with the definition $\bs{U}^\perp(\omega,\bp,\bp'')\equiv\bs{U}^\perp_{\omega\bp\bp''}(0,0)$.
In coordinate representation, the fourth-rank tensor $\bs{\alpha}^\perp(\omega,\bp)$ is given by
\begin{eqnarray}
\alpha_{ij,kl}^\perp(\omega,\bp)=-\frac{c^2}{\omega^2}\left[
\text{Re}\Sigma^\perp(\omega,p)P_{ik}(\bp)P_{jl}(\bp)+
\intdtwo{\bp'}\text{Re}\Sigma^\perp(\omega,p')U_{ij,kl}^\perp(\omega,\bp',\bp)
\right].
\label{Aijkl}
\end{eqnarray}
In order to evaluate $\lambda(\bq, \Omega)$, we substitute the solution (\ref{diff_sol}) for $\bs\Phi^\perp_{\omega\bp\bp'}$ in Eq. (\ref{kinetic_eq_tr}) using Eqs. (\ref{expansion_phi}) and (\ref{expansion_lambda}) and proceed in two steps. First, we take the limit $\bq\to 0$ in Eq. (\ref{kinetic_eq_tr}), integrate over $\bp$ and $\bp'$ and trace over tensor components. This gives
\begin{equation}
\lambda(0,\Omega)=-\frac{i\Omega\omega a}{c^2},
\label{lambda_Omega}
\end{equation}
where
\begin{equation}
a=\left[2\intdtwo{\bp}\text{Im}\overline{G}^\perp(\omega,p)\right]^{-1}\times\intdtwo{\bp}\text{Im}\overline{G}^\perp(\omega,p)\alpha^\perp(\omega,\bp),
\label{adef2}
\end{equation}
with $\alpha^\perp(\omega,\bp)=\bs{P}(\bp)\cdot\bs{\alpha}^\perp(\omega,\bp)\cdot\bs{P}(\bp)\equiv P_{ij}(\bp)\alpha_{ij,kl}^\perp(\omega,\bp)P_{kl}(\bp)$. Second, we take the limit $\Omega\to 0$ in Eq. (\ref{kinetic_eq_tr}), sum over $\bp$ and $\bp'$ and trace over tensor components. This leads to
\begin{equation}
\delta\lambda(\bq,\Omega)=\left[-2\intdtwo{\bp}\text{Im}\overline{G}^\perp(\omega, p)\right]^{-1}\times\frac{\bq^2}{3}\intdtwo{\bp}p_m J_{ii,m}(\omega,\bp).
\label{lambda_q}
\end{equation}
Inserting the results (\ref{expansion_phi}), (\ref{lambda_Omega}) and (\ref{lambda_q}) into Eq. (\ref{diff_sol}), we infer
\begin{equation}
\Phi^{\perp ij,kl}_{\omega\bp\bp'}(\bq,\Omega)\sim 
\frac{\text{Im}\overline{G}^\perp(\omega,p)\text{Im}\overline{G}^\perp(\omega,p')}
{-i\Omega\omega/c^2(1+a)+\delta\lambda(\bq,\Omega)}
P_{ik}(\bp)P_{jl}(\bp'),
\end{equation}
where we have dropped the $\bq$-dependent terms in the numerator. To obtain the light intensity (\ref{intensity_Phi}), we finally contract this result with the polarization vectors $\bs\epsilon_\text{in}$ and $\bs\epsilon_\text{out}$ and integrate over $\bp$. This leads to
\begin{equation}
\overline{I}_\omega(\bp', \br,t)\sim\intdone{\Omega}\intdtwo{\bq}\frac{e^{i\bq\cdot\br-i\Omega t}}{-i\Omega+D\bq^2}A(\omega,p')[1-(\hat{\bp}'\cdot\bs\epsilon_\text{out})^2],
\end{equation}
which is Eq. (\ref{diff_intensity}) of the main text (with $\bp'$ and $\bs\epsilon_\text{out}$ relabeled $\bp$ and $\bs\epsilon$, respectively). We have here introduced the spectral function
\begin{equation}
A(\omega,p')=-\frac{2\omega}{\pi c^2}\text{Im}\overline{G}^\perp(\omega,p')=-\frac{2\omega}{\pi c^2}\frac{\text{Im}\Sigma^\perp(\omega,p')}{[\omega^2/c^2-\text{Re}\Sigma^\perp(\omega,p')-\bp'^2]^2+[\text{Im}\Sigma^\perp(\omega,p')]^2},
\label{Kubo_Greenwood}
\end{equation}
which to first order in $\eta$ leads to Eq. (\ref{spectral_f}) of the main text, with $v_\varphi$ given by Eq. (\ref{vphi}) and $\ell_s$ related to $\Sigma^\perp(\omega, p')$ through Eq. (\ref{ls_def}).
The diffusion coefficient $D$ is given by 
\begin{equation}
D=\left[-2\intdtwo{\bp}\text{Im}\overline{G}^\perp(p)\right]^{-1}\times\frac{c^2}{3\omega(1+a)}\intdtwo{\bp}p_m J_{ii,m}(\omega,\bp),
\label{Kubo_Greenwood}
\end{equation}
which has the form of a Kubo-Greenwood formula \cite{Barabanenkov91, Barabanenkov95a}.

\subsection{Transport mean free path and energy transport velocity}

At this stage, the current tensor $\bs{J}$ in Eq. (\ref{Kubo_Greenwood}) is still unknown. A self-consistent equation for $\bs{J}$ can be found by inserting Eq. (\ref{diff_sol}) into Eq. (\ref{kinetic_eq_tr}) evaluated at $\Omega=0$. This gives:
\begin{equation}
J_{ij,m}(\omega,\bp)=p_m|\overline{G}^\perp(\omega,p)|^2P_{ij}(\bp)+|\overline{G}^\perp(\omega,p)|^2\intdtwo{\bp'}U^\perp_{ij,kl}(\bp,\bp')J_{kl,m}(\omega,\bp').
\label{Jdef}
\end{equation}
We explicitly solve this equation by making use of an on-shell approximation, which turns out to be exact at order $\eta^2$ \cite{vanTiggelen94}. The latter consists in evaluating $U^\perp_{ij,kl}(\bp,\bp')$ at $p\simeq p'\simeq \omega/v_\varphi$, using that $|\overline{G}^\perp(\omega,p)|^2$ is a narrow function of $p$, peaked around $p=\omega/v_\varphi$. After iteration of Eq. (\ref{Jdef}), this allows us to write
\begin{equation}
J_{ij,m}(\omega,\bp)=p_m|\overline{G}^\perp(\omega,p)|^2P_{ij}(\bp)\times\left\{1-\frac{1}{2}\left[\intdtwo{\bp}|\overline{G}^\perp(\omega,p)|^2\right]\times
\langle \hat{\bp}\cdot \hat{\bp}' U^\perp(\omega,k\hat{\bp},k\hat{\bp}')
\rangle_{\hat{\bp}'}\right\}^{-1},
\label{Jdef_iterated}
\end{equation}
where we have introduced $U^\perp(\omega,\bp,\bp')=\bs{P}(\bp)\cdot\bs{U}^\perp(\omega,\bp,\bp')\cdot\bs{P}(\bp')=\bs{P}(\bp)\cdot\bs{U}(\omega,\bp,\bp')\cdot\bs{P}(\bp')$.
Eq. (\ref{Jdef_iterated}) is further simplified by invoking the Ward identity for $\Omega=0$ and $\bq=0$:
\begin{equation}
\text{Im}\Sigma^\perp(\omega,p)P_{ik}(\bp)P_{jl}(\bp)=\intdtwo{\bp'}\text{Im}\overline{G}^\perp(p')U^\perp_{ij,kl}(\omega,\bp,\bp'),
\label{Ward_Omega_zero}
\end{equation}
which after use of the on-shell approximation and trace over tensor components leads to $\intdtwo{\bp}|G^\perp(\omega,p)|^2=2/\langle U^\perp(\omega,k\hat{\bp},k\hat{\bp}')\rangle_{\hat{\bp}'}$. Inserting this result into Eq. (\ref{Jdef}), we obtain
\begin{equation}
J_{ij,m}(\omega,\bp)=\frac{2p_m|\overline{G}^\perp(\omega,p)|^2P_{ij}(\bp)}{\intdtwo{\bp}|G^\perp(\omega,p)|^2}
\times\frac{1}{
\langle U^\perp(\omega,k\hat{\bp},k\hat{\bp}')(1-\hat{\bp}\cdot \hat{\bp}')
\rangle_{\hat{\bp}'}}.
\label{Jfinal}
\end{equation}
We finally insert this relation into the Kubo formula (\ref{Kubo_Greenwood}) and again use the on-shell approximation to carry out the integrals involving $|\overline{G}^\perp(\omega,p)|^2$. This gives
\begin{equation}
D=\frac{c^2}{3v_\varphi(1+a)}\frac{8\pi}{
\langle U^\perp(\omega,k\hat{\bp},k\hat{\bp}')(1-\hat{\bp}\cdot \hat{\bp}')
\rangle_{\hat{\bp}'}},
\end{equation}
which is Eq. (\ref{diff_coefficient}) of the main text, with $\ell^*$ given by Eq. (\ref{lstar_def}). The formulation (\ref{adef}) of $a$ finally follows from Eqs. (\ref{adef2}) and (\ref{Aijkl}) combined with the Ward identity (\ref{Ward_Omega_zero}).

\section{Transverse part of irreducible vertices}
\label{appendix_b}

In this appendix, we give the explicit expressions of the transverse components of $\bs\Sigma^{(2)}$ and $\bs U^{(2)}$ involved in the calculation of the stored electromagnetic energy $a$ [Eq. (\ref{deltaa})] and of the transport mean free path [Eq. (\ref{delta_sigma_vec})]. 

$\Sigma^{\text{(2,a)}\perp}(\omega)$ and $\Sigma^{\text{(2,b)}\perp}(\omega,k)$ follow straightforwardly from the decomposition (\ref{G0_eq}) of $\bs{G}_0$:
\begin{eqnarray}
\Sigma^{\text{(2,a)}\perp}(\omega)=
n^2 t^3\intdr{\br}\,
\left[\frac{2}{3}\frac{G^{\perp2}_0(r)}{1-t^2G^{\perp2}_0(r)}+\frac{1}{3}\frac{G^{\parallel2}_0(r)}{1-t^2G^{\parallel2}_0(r)}
\right]
\label{Sigma_2a_perp}
\end{eqnarray}
and
\begin{eqnarray}
\Sigma^{\text{(2,b)}\perp}(\omega,k)=
n^2 t^4\intdr{\br}\,
\left\{\left[j_0(kr)-\frac{j_1(kr)}{kr}\right]\frac{G^{\perp3}_0(r)}{1-t^2G^{\perp2}_0(r)}+\frac{j_1(kr)}{kr}\frac{G^{\parallel3}_0(r)}{1-t^2G^{\parallel2}_0(r)}
\right\}-\frac{n^2 t^2}{3k^2},
\label{Sigma_2b_perp}
\end{eqnarray}
where $G_0^\perp(r)=[-1+1/(ikr)+1/(kr)^2]e^{ik r}/(4\pi r)$ and $G_0^\parallel(r)=-2[1/(ikr)+1/(kr)^2]e^{ik r}/(4\pi r)$. $j_0$ and $j_1$ are spherical Bessel functions. The last term in Eq. (\ref{Sigma_2b_perp}) stems from the singular part of the Green tensor (\ref{G0_eq}), which we have explicitly separated from $G_0^\perp(r)$ and $G_0^\parallel(r)$. 

We then consider the two angular averages in Eq. (\ref{sigma_star_tot}). Their evaluation requires first to expand the ratio of tensors in the integrand of Eqs. (\ref{U2a}) and (\ref{U2b}) over a basis of orthogonal eigentensors, and then to carry out the angular integrals over the directions of $\br$ and $\bp'$. After a tedious calculation we find
\begin{eqnarray}
\frac{\langle\hat{\bp}\cdot\hat{\bp}' U^{(2,a)\perp}(\omega, k\hat\bp,k\hat\bp')\rangle_{\hat\bp'}}{8\pi}&=&
\intdr{\br}\, \frac{n^2|t|^4}{4\pi}\bigg\{
A(r)\left|\frac{G^\perp_0(r)}{1-t^2G^{\perp2}_0(r)}\right|^2
+B(r)\left|\frac{G^{\perp}_0(r)}{1-t^2G^{\perp2}_0(r)}-\frac{G^{\parallel}_0(r)}{1-t^2G^{\parallel2}_0(r)}\right|^2\nonumber\\
&&+2C(r)\text{Re}\frac{G^{\perp}_0(r)}{1-t^2G^{\perp2}_0(r)}
\left[\frac{G^{\parallel}_0(r)}{1-t^2G^{\parallel2}_0(r)}-\frac{G^{\perp}_0(r)}{1-t^2G^{\perp2}_0(r)}\right]^*\bigg\}
\end{eqnarray}
and
\begin{eqnarray}
\frac{\langle\hat{\bp}\cdot\hat{\bp}' U^{(2,b)\perp}(\omega, k\hat\bp,k\hat\bp')\rangle_{\hat\bp'}}{8\pi}&=&
-\intdr{\br}\, \frac{n^2|t|^2}{4\pi}\bigg\{
A(r)\left[\left|\frac{1}{1-t^2G^{\perp2}_0(r)}\right|^2-1\right]
+2B(r)\left|\frac{G^{\perp}_0(r)}{1-t^2G^{\perp2}_0(r)}-\frac{G^{\parallel}_0(r)}{1-t^2G^{\parallel2}_0(r)}\right|^2\nonumber\\
&&+2C(r)\text{Re}\frac{1}{1-t^2G^{\perp2}_0(r)}
\left[\frac{1}{1-t^2G^{\parallel2}_0(r)}-\frac{1}{1-t^2G^{\perp2}_0(r)}\right]^*\bigg\},
\end{eqnarray}
where $A(r)=j_1^2(kr)+[j_2(kr)j_3(kr)-j_1(kr)j_2(kr)]/(kr)$, $B(r)=2j_2^2(kr)^2/(kr)^2$ and $C(r)=3j_2^2(kr)^2/(kr)^2$.\newline
\end{widetext}


\begin{thebibliography}{99}
\bibitem{footnote1}
Note that this statement cannot be generalized for $\bs{\Sigma}^{(3)}$, whose calculation requires to consider diagrams which contain infinitely many scatterers.
\bibitem{footnote2}
Strictly speaking, $V_\text{DD}(r)$ is the induced dipole-dipole potential per unit of energy density of the radiation.
\bibitem{footnote3}
Note in passing that this estimation also confirms that, at low densities, transport via $\overline{G}^\parallel$ is completely negligible as compared to the binary IDDC mechanism discussed in the present paper.
\end{thebibliography}
\end{document}